\pdfoutput=1
\documentclass[a4paper,fleqn,usenatbib,umoline]{mnras}

\usepackage{amsmath}  
\usepackage{amssymb}  
\usepackage{graphicx} 
\usepackage{txfonts}
\usepackage{ae,aecompl}
\usepackage{color}
\usepackage[multidot]{grffile}
\usepackage[capitalise]{cleveref}
\crefname{eq:}{Eq.}{Eqs.}
\crefname{fig:}{Fig}{Figs}
\usepackage{subfig}
\usepackage{umoline}
\usepackage{siunitx}
\usepackage{upgreek}
\usepackage[normalem]{ulem}
\usepackage{wasysym}
\usepackage{booktabs}
\usepackage{threeparttable}  

\newcommand{\be}{\begin{equation}}
\newcommand{\ee}{\end{equation}}
\newcommand{\bvec}[1]{\boldsymbol{#1}}
\newcommand{\tpartial}[1]{\frac{\partial\, #1}{\partial t}}
\newcommand{\cs}{c_\mathrm{s}}

\newcommand{\rhoc}{\rho_\mathrm{c}}

\defcitealias{Rad2018}{H18}

\title[Gravitational instability of molecular clouds]{Gravitational instability of non-isothermal filamentary molecular clouds, in presence of external pressure}

\author[Motiei et al.]{
Mohammad Mahdi Motiei,$^{1}$\thanks{E-mail: motiei@um.ac.ir (MMM); hosseinirad@um.ac.ir (MH) abbassi@um.ac.ir (SA)}
Mohammad Hosseinirad$^{1,2}$
and Shahram Abbassi$^{1}$
\\
 $^{1}$Department of Physics, School of Sciences, Ferdowsi University of Mashhad, Mashhad, PO Box 91775-1436, Iran\\
 $^{2}$School of Astronomy, Institute for Research in Fundamental Sciences (IPM), PO Box 19395-5531, Tehran, Iran
}
\date{Accepted . Received ; in original form }

\pubyear{2021}

\begin{document}
\label{firstpage}
\pagerange{\pageref{firstpage}--\pageref{lastpage}}
\maketitle
\begin{abstract}
Filamentary molecular clouds are omnipresent in the cold interstellar medium.  
Observational evidences show that the non-isothermal equations of state describe the filaments properties better than the isothermal one. In this paper we use the logatropic and the polytropic equations of state to study the gravitational instability of the pressure-confined filaments in presence of a uniform axial magnetic field. To fully explore the parameter space we carry out very large surveys of stability analysis that cover filaments with different radii in various magnetic fields. Our results show that for all the equations of state the instability of thinner filaments is more sensitive to the magnetic field variations than the thicker ones. Moreover, for all the equations of state, an intermediate magnetic field can entirely stabilize the thinner filaments. Albeit for the thicker ones this effect is suppressed for the magnetic field stronger than $B\simeq 70$ $\upmu$G.
\end{abstract}


\section{Introduction}
The cold interstellar molecular gas in the Galaxy has been revealed to have filamentary structures of parsec-scale (0.5\,-\,100 pc) \citep[e.g.][]{Schneider1979,Bally1987,Goldsmith2008,Andre2017}, regularly harbouring clumps and dense cores \citep[e.g.][]{2007ARA&A..45..339B,2010A&A...520A.102M,2010ApJ...719L.185J,2011ApJ...735...64W,2012A&A...542A.101M,2014MNRAS.439.3275W,2016MNRAS.456.2041C,2016MNRAS.463..146H,2016ApJS..226....9W,2016A&A...592A..21F,Kainulainen2016}. The filamentary molecular clouds (MCs) unveiled by unprecedented images of \emph{Herschel Space Observatory} \citep{Pilbratt2010}, represent a common width of $\sim 0.1$ pc \citep[\citet{Arzoumanian2019}, see also][for a recent debate on the existence of such a universal width]{Panopoulou2017,Andre2017,Roy2019} at least in the nearby Gould belt, while they extend over a wide range in column density. The filamentary MCs, are identified both in non-star-forming \citep{Menshchikov2010,Miville-Deschnes2010,Ward-Thompson2010} and star-forming \citep{Konyves2010,Bontemps2010} regions which emphasizes their importance to better understand the theory of star formation \citep{Andre2014}.

In the filamentary picture of formation of stars, the large-scale turbulent flows are assembled into a network of filaments due to the supersonic shocks \citep[e.g.][]{Klessen1998,McKee-2007,Dib2007,Padoan2014,Pudritz2013} or combination with the magnetic field which is most probably perpendicular to the filaments \citep[e.g.][]{Nakamura2008,Chen2014,Inutsuka2015,Federrath2016,Klassen2017,Li2019}. It is also possible that the global collapse of the parent cloud under its self-gravity, governs the formation process \citep[e.g.][]{Burkert2004,Hartmann2007,Vazquez-Semadeni2007,Gomez2014,Wareing2016,Camacho2016}. If the filaments are gravitationally unstable, they will fragment onto the cores and finally form clusters of stars \citep{Lada2003} provided that the conditions for the subsequent fragmentation are met.

Filaments have been subject to many investigations since almost the mid-point of the twentieth century, when the groundbreaking work by \cite{Chandra53} showed that a poloidal magnetic field is able to completely stabilize a very long uniform incompressible cylinder of gas. Ten years later, \citet{Stod63} derived the magnetostatic equilibrium of an isothermal gas cylinder threaded by a longitudinal magnetic field proportional to the square root of its initial density. Physical explanation of the filamentary clouds was interesting enough to encourage other authors for more detailed theoretical investigations \citep[see e.g.][]{Larson1985,Nagasawa87,Inutsuka1992,Nakamura1993,Matsumoto94,Gehman1,Gehman2,Inutsuka1997,Fischera2012,Freundlich2014,Hanawa2015,Sadhukhan2016,Rad2017,Hanawa2017,Hanawa2019}. In addition, various simulations of the cylindrical geometry have been performed for more realistic studies in non-linear regime \citep[see e.g.][]{Steinacker2016,Gritschneder2017,Heigl2016,Heigl2018,Heigl2020,Ntormousi2019,Clarke2017,Clarke2020}.

Filaments are seldom found in isolation but under the external pressure of the ambient medium \citep{Fischera2012,Fischera2012b}. \citet{Nagasawa87} performed global linear stability analysis for an infinitely long isothermal magnetized filament. \citet{Nagasawa87} showed that both a non-confined and a pressure-confined filament are gravitationally unstable for a specific range of wavelengths. More specifically, \citet{Nagasawa87} found that a poloidal magnetic field can increase the stability of a filament and interestingly entirely stabilize it if the filament is thin enough. The linear stability analysis of self-gravitating objects under the effect of external pressure in other environments, have been also the matter of many studies (see e.g. \citealt[][for sheet-like gas layers]{Miyama1987a,Nagai1998,Durrive2019} and also
\citealt[][for gas disks]{Chou2000,Lee2007,Kim2012}). Recently, \cite{anathpindika2020filament} have reported the external pressure could affect the peak central density, the column density, the morphology and the star formation of the filaments.
	
Based on the column density maps extracted from \emph{Herschel} images \citep{Arzoumanian2011,Juvela2012,Palmeirim2013}, the radial density profiles of the filaments in Gould belt, could not be properly described by a simple isothermal model \citep{Stod63,Ostriker64}, but instead they are best fitted by the softer (i.e. profiles which are shallower at distances away from the centre) polytropic models with indices $\gamma_p < 1$ \citep{Palmeirim2013,Toci2015}. Moreover, for the filaments in the IC5146 region, a modification to the simple isothermal model that supposes a very long subcritical pressure-confined cylinder with different masses per unit length, can account for this problem (\citealt{Fischera2012}; see also \citealt{Heitsch2013a} for a similar but accreting model), which is also the case for a near-critical cylinder wrapped by a helical magnetic field \citep{FP2000II}. In addition, polytropic filaments with indices less than but near the unity that are undergoing gravitational collapse, have also shallow density profiles \citep{Kawachi1998,Nakamura1999,Shadmehri2005}.

In a recent paper, \citet[][hereafter H18]{Rad2018} carried out a similar analysis to the work that had been done by \citet{Nagasawa87}, but for the aforementioned polytropic equation of state (PEOS) as well as the logatropic equation of state (LEOS) \citep{LizanoShu1989}. They used the non-ideal magnetohydrodynamic (MHD) framework for a filament threaded by a poloidal magnetic field in the absence of the external pressure. They found that without the effect of magnetic field, filaments with these two softer types of equations of state (EOSs) are more susceptible to the gravitational instability than a filament with the isothermal EOS (IEOS). More specifically, they realized that while the gravitational instability in a moderate magnetized filament is generally sensitive to the type of EOS, the instability is suppressed in the strongly magnetized one, regardless of its EOS type. Here, we aim to elucidate how a pressure-confined filament with the LEOS or the PEOS responds to the linear perturbations, therefore combining the study by \cite{Nagasawa87} and \citetalias{Rad2018}, albeit in the ideal MHD for simplicity. We will investigate this problem in the non-ideal MHD in a forthcoming paper.

The outline of this paper is as follows.
In Section \ref{sec:MHD}, 
we explain the ideal MHD equations considering self-gravity. Section \ref{sec:EOS} introduces the non-isothermal EOSs we use in this paper. The equilibrium state and perturbations are described in sections \ref{sec:unperturbed} and \ref{sec:perturbed}, respectively. Section \ref{sec:boundary} deals with the boundary conditions. Computation method is given in section \ref{sec:method}. Section \ref{sec:results} and Section \ref{sec:conclusion} contain the results and conclusions of this investigation.

\section{Basics equations and formulations}
\subsection{Ideal MHD equations considering self-gravity}\label{sec:MHD}
We consider an infinitely long cylinder of gas with finite radius as the filament. The filament is threaded by a uniform magnetic field parallel to its long axis so $\bvec{B}=(0, 0, B_{z})$ which does not affect the unperturbed structure. Our set of equations include the equation of motion \eqref{eqforce}, the induction equation \eqref{eqinduction}, the continuity equation \eqref{eqcont} and the Poisson's equation \eqref{eqpoisson} as
\begin{equation}
\rho\tpartial{\bvec{u}} + \rho\left(\bvec{u}\cdot\nabla\right) \bvec{u}
+ \nabla p + \rho\nabla\psi
- \frac{1}{4\pi} \left(\nabla\times\bvec{B}\right)\times\bvec{B} = 0,
\label{eqforce}
\end{equation}
\begin{equation}
\tpartial{\bvec{B}} + \nabla\times\left(\bvec{B}\times\bvec{u}\right) = 0,
\label{eqinduction}
\end{equation}
\begin{equation}
\tpartial{\rho} + \nabla\cdot\left(\rho\bvec{u}\right) = 0,
\label{eqcont}
\end{equation}
\begin{equation}
\nabla^{2}\psi = 4 \pi G \rho.
\label{eqpoisson}
\end{equation}
In equations \eqref{eqforce} to \eqref{eqpoisson}, $\rho$, $\bvec{u}$, $p$, $\bvec{B}$ and $\psi$ indicate the neutral gas density, the velocity of the fluid, the gas pressure, the magnetic field strength and the gravitational potential, respectively.
\subsection{Equations of state}\label{sec:EOS}
Equations \eqref{eqforce} to \eqref{eqpoisson} must be accompanied by a relation between the pressure and the density in order to be complete. There are many studies that use the IEOS (see e.g. \citealt{Nagasawa87, Inutsuka1992, Fischera2012b, Heigl2016, Rad2017}).  We consider three different types of non-isothermal EOSs. In the following we describe them briefly.
\subsubsection{Gehman equation of state (GEOS)}
By applying a non-isothermal barotropic equation of state, \citet{Gehman1,Gehman2} studied the observed turbulence effect in MCs. They added a term to the IEOS in order to model this effect. This EOS is softer than the IEOS and there is theoretical and empirical support for using that \citep[][and \citealp{LizanoShu1989}]{Gehman1,Gehman2}. They proposed the GEOS form as
\begin{equation}
p = c_{\rm s}^{2}\rho + \mathcal{P}_{0} \ln(\rho/\rho_{\rm c}).
\label{eqGEOS}
\end{equation}
In this equation $c_{\rm s}$ and $\rho_{\rm c}$  are the isothermal sound speed and the density at the filament axis, respectively. $\mathcal{P}_{0}$ is an empirical constant which its value changes between 10 and 70 picodynes cm$^{-2}$ \citep{Gehman2}.
\subsubsection{McLaughlin \& Pudritz eqaution of state (MPEOS)} 
 \cite{MP96} considered a pure logarithmic EOS as
\begin{equation}
p = p_{\rm c}[1 + A \ln(\rho/\rho_{\rm c})]
\label{eqMPEOS},
\end{equation}
where $p_{\rm c}$ and $\rho_{\rm c}$ are the pressure and the density along the filament axis, respectively and $A$ is an empirical constant about 0.2 for molecular cloud cores. They claimed that this EOS is the simplest and the most successful model that contains the important properties of the giant MCs and their internal structures such as cores. Also \cite{FP2000I} found that this logatropic model is in agreement with the existing data and although it was based on the core data, they used the same value of $A$ for the filamentary clouds. This EOS is the softest one among the EOSs we use in this text.
\subsubsection{Polytropic equation of state (PEOS)}\label{sec:PEOS}
\cite{Palmeirim2013} argued that the structure of B211 filament in the Taurus MC is well described by a polytropic cylindrical filament with an EOS as  $p\propto\rho^{\gamma}$  where $\gamma = 0.97\pm 0.01.$ \cite{Toci2015}
analysed the observational properties of the filamentary clouds in the cylindrical symmetry with the PEOS and the polytropic exponent $1/3 \apprle \gamma_{\rm p} \apprle 2/3$ (the polytropic indices $-3 \apprle n \apprle-1.5$) where $\gamma_{\rm p} = 1+1/n.$ In a more general way $-\infty<n<-1$ \citep{Viala1974,Maloney1988}.
In this paper we use the PEOS with negative index as
\begin{equation}
p=p_{\rm c}(\rho/\rhoc)^{1+1/n}
\label{eqPEOS},
\end{equation}
where $p_{\rm c}$ and $\rho_{\rm c}$ are the same as in equation \eqref{eqMPEOS}. The dimensionless forms of equations \eqref{eqGEOS} to \eqref{eqPEOS} are available in appendix \ref{sec:dimensionless}.
\subsection{Equilibrium state}\label{sec:unperturbed}
We use the cylindrical coordinates $(r, \phi, z)$ by assuming the filament centre at the origin. The filament is very long and its radius is confined. The initial magnetic field $\bvec{B_0} = B_0\,\hat{z}$ is uniform and has not any effect on the equilibrium of the filament. Solving a combination of equations \eqref{eqforce} and \eqref{eqpoisson} gives us the density profile at the equilibrium state. For a filament with the IEOS we have
\begin{equation}
 \rho(r)=\rhoc (1+\dfrac{r^2}{8H^2})^{-2},
 \label{eq:rho}
\end{equation}
where $H$ is the radial scale length \citep{Stod63,Ostriker64} as
\begin{equation}
 H = \dfrac{c_{\rm s}}{\sqrt{4\pi G \rhoc}}
 \label{eq:H}
\end{equation}
where $G$ is the gravitational constant. For a typical MC with the central density of 4$\times$10$^{-20}$ g cm$^{-3}$ and the thermal sound speed of 0.2 km s$^{-1}$, $H$ $\approx 0.035$ pc. The magnetic field strength $B$ is in the unit of $(4 \pi \rhoc)^{1/2}c_{\rm s}$ and considering the mentioned values for $\rhoc$ and $c_{\rm s}$, $B = 1$ is equivalent with $B \simeq 14.2$ $\upmu$G. See Appendix \ref{sec:dimensionless} for more details.
Since there are no analytical solutions for the non-isothermal EOSs presented in $\S$ \ref{sec:EOS}, we obtain them numerically (\citetalias{Rad2018}). We set the dimensionless turbulence parameter $\kappa = 0.1, 0.2, 0.5, 1$ in the GEOS where $\kappa = \mathcal{P}_{\rm 0}/(c_{\rm s} ^2 \rhoc)$, the dimensionless parameter $A = 0.2$ in the MPEOS and the polytropic indices $n = -1.5, -2, -3, -4$ in the PEOS.
\subsection{Perturbations}\label{sec:perturbed}
By applying a small perturbation of $\delta r$ to the surface of the filament \citep{Gehman2}, the perturbed version of equations \eqref{eqforce} to \eqref{eqpoisson} in dimensionless form to the first order gives us
\be
\rho_{0}\tpartial{\bvec{u}_{1}} + \nabla p_{1} + \rho_{0}\nabla\psi_{1}
+ \rho_{1}\nabla\psi_{0} - \left(\nabla\times\bvec{B}_1\right)\times\bvec{B}_0 = 0,
\label{eq:mom}
\ee
\be
\tpartial{\bvec{B}_{1}}
+ \nabla\times\left(\bvec{B}_{0}\times\bvec{u}_{1}\right) = 0
\label{eq:ind1},
\ee
\be
\tpartial{\rho_{1}} + \nabla\rho_0\cdotp\bvec{u}_1 + \rho_0\nabla\cdotp\bvec{u}_1=0 
\label{eq:cont1},
\ee
\be
\nabla^{2}\psi_{1} = \rho_{1}
\label{eq:lpsi1}.
\ee
In these equations, the subscript 0 indicates the unperturbed parameters and the subscript 1 shows the perturbed quantities. Since our EOSs are barotropic, we can linearize them as
\be p_1=\dfrac{d p}{d\rho}(\rho_0)\rho_1 \equiv p^{\prime}(\rho_0)\rho_1
\label{eq:p1}.
\ee
On the other hand the density, the velocity, the magnetic field and the gravitational potential are as
\be
\rho_{1}(\bvec{x}, t) = \mathbb{R} \left[ f(r) \exp{(ikz-i\omega t)} \right],
\label{eq:rho1}
\ee
\be
\bvec{u}_{1}(\bvec{x}, t) = \mathbb{R} \left[ \bvec{v}(r) \exp{(ikz-i\omega t)} \right],
\label{eq:u1}
\ee
\be
\bvec{B}_{1}(\bvec{x}, t) = \mathbb{R} \left[ \bvec{b}(r) \exp{(ikz-i\omega t)} \right],
\label{eq:B1}
\ee
\be
\psi_{1}(\bvec{x}, t) = \mathbb{R} \left[ \phi(r) \exp{(ikz-i\omega t)} \right],
\label{eq:psi1}
\ee
where $\mathbb{R}$ refers to the real part, $k$ is the wave number (along the filament axis) and $\omega$ denotes to the angular frequency. $f(r)$, $\bvec{v}(r)$, $\bvec{b}(r)$ and $\phi(r)$ are the amplitudes of the perturbations. We use these forms and equations \eqref{eq:mom} to \eqref{eq:cont1} in order to get the linearised dimensionless forms by considering $w$ = $i\omega v_{r}$ in the following equations
\be
\rho_{0} (\frac{d}{dr} + \frac{1}{r}) w + (\omega^{2} - k^{2}P') f - k^{2} \rho_{0} \phi + w \frac{d\rho_{0}}{dr}=0,
\label{eq:rhoddr}
\ee
\be
\bigg[
P' + \frac{B_{0}^{2}}{\rho_{0}}
\bigg(
1-\frac{k^{2}}{\omega^{2}} P'
\bigg)
\bigg]
\frac{df}{dr}
+ A_{1}f
+
\bigg(
\rho_{0} - \frac{k^{2}}{\omega} B_{0}^{2}
\bigg)
\frac{d\phi}{dr}
+ A_{2}\phi + A_{3} w = 0,
\label{eq30gehman}
\ee
\be
(\frac{d}{dr} + \frac{1}{r}) \frac{d\phi}{dr} - k^{2}\phi - f = 0,
\label{eq:ddr}
\ee
where
\be
A_{1} = -\frac{k^{2}B_{0}^{2}}{\omega^{2} \rho_{0}}
\bigg(
P^{\prime\prime} - \frac{2P'}{\rho_{0}}
\bigg)
\frac{d\rho_{0}}{dr}
+
\bigg(
P^{\prime\prime} - \frac{2B_{0}^{2}}{\rho_{0}^{2}}
\bigg)
\frac{d\rho_{0}}{dr}
+ \frac{d\psi}{dr},
\ee
\be
A_{2} = \frac{k^{2}B_{0}^{2}}{\omega^{2} \rho_{0}} \frac{d\rho_{0}}{dr},
\ee
\be
A_{3} = \frac{B_{0}^{2}}{\omega^{2}}
\bigg[
k^{2} + \frac{1}{\rho_{0}} \frac{d^{2}\rho_{0}}{dr^{2}} - 2
\bigg(
\frac{1}{\rho_{0}} \frac{d\rho_{0}}{dr}
\bigg)^{2}
- \frac{1}{r\rho_{0}} \frac{d\rho_{0}}{dr}
\bigg]
- \rho_{0}.
\ee
\subsection{Boundary conditions}\label{sec:boundary}
In this section we closely follow \cite{Nagasawa87} approach for obtaining boundary conditions. The external pressure confines the filament to the finite radius $R$. The perturbed surface of this filament will have the radius
\be
r=R+\delta r \exp (ikz - i \omega t).
\ee
On the deformed surface of the filament, the $r$-component of the velocity is then defined as 
\be
v_{r} (R) = -i \omega \delta r
\label{urR}.
\ee
On the other hand the pressure on the boundary must be equal to the external pressure. This leads to
\begin{equation}
p_{0}(R)= \begin{cases}
\rho_{0}(R) + \kappa \ln[\rho_{0}(R)] & \text{(GEOS)}, \\[2\jot]
1 + A \ln[\rho_{0}(R)] & \text{(MPEOS)}, \\[2\jot]
\rho_{0}(R)^{1+1/n} & \text{(PEOS)},
\end{cases}
\label{p_0}
\end{equation}
in the dimensionless form and to the first order of the perturbation we will have
\be
\frac{dp_{0}}{dr} \bigg|_{R} \delta r + p_{1}(R) + B_{0} B_{1z} (R) = B_{0} B_{1z}^{ext} (R),
\label{dp0_dr}
\ee
It is necessary for the gravitational potential and its radial derivative to be continuous on the border. So
\be
\psi_{1}(R) = \psi_{1}^{ext}(R)
\label{psi1_psi1ext},
\ee
\be
\frac{d\psi_{1}}{dr} \bigg|_{R} + \rho_{0} (R) \delta r = \frac{d\psi_{1}^{ext}}{dr} \bigg|_{R}
\label{dpsi1_dpsi1ext}.
\ee
We consider a very hot and low density environment out of the filament $(r>R)$. We solve Laplace's equation for exterior gravitational potential in cylindrical coordinates and write the solution with the modified Bessel function of the second type and order $m$ ($K_{m}$). Because we only investigate axisymmetric and unstable modes, we use $m = 0$ order in the modified Bessel function, so we restrict ourselves to the axisymmetric mode ($m = 0$). This is because it was shown by \citep{Nagasawa87} that non-axisymmetric modes ($m \geqslant 1$) are stable against perturbation. This will let us to recast equation \eqref{dpsi1_dpsi1ext} into
\be
\frac{d \psi_{1}}{dr} \bigg|_{R} + \rho_{0} (R) \delta r = -k  \frac{K_{1} (kR)}{K_{0}(kR)}  \psi_{1} (R)
\label{dpsi1_dr}.
\ee
Furthermore, we consider that there is no electric current outside the filament. So, $B_{1z}$ should be continuous on the boundary. Considering equations \eqref{p_0} and \eqref{dp0_dr} we will have for the GEOS
\be
\begin{aligned}
\bigg(1+\frac{\kappa}{\rho_{0}}\bigg) \bigg(\rho_{1} + \frac{d\rho_{0}}{dr}  \delta r\bigg) \bigg|_{R} + B_{0} B_{1z} (R) ={}& \\ -B_{0} \frac{iK_{0}(kR)}{K_{1}(kR)} B_{1r} (R),
\end{aligned}
\label{dp0_GEOS}
\ee
for the MPEOS
\be
\begin{aligned}
\bigg(\frac{A}{\rho_{0}} \bigg) \bigg( \rho_{1}+ \frac{d\rho_{0}}{dr}  \delta r \bigg)\bigg|_{R} + B_{0} B_{1z} (R) ={}& \\ -B_{0} \frac{iK_{0}(kR)}{K_{1}(kR)} B_{1r} (R),
\end{aligned}
\label{dp0_MPEOS}
\ee
and for the PEOS
\be
\begin{aligned}
	\bigg(1+\frac{1}{n}\bigg) \rho_{0}^{1/n} \bigg(\rho_{1} + \frac{d\rho_{0}}{dr} \delta r \bigg)\bigg|_{R} + B_{0} B_{1z} (R) ={}& \\ -B_{0} \frac{iK_{0}(kR)}{K_{1}(kR)} B_{1r} (R).
\end{aligned}
\label{dp0_PEOS}
\ee
We set the boundary conditions along the filament axis $(r=0)$ as
\be
f=1,\quad \frac{d\phi}{dr}=0,\quad w=0.
\label{bc_r_0}
\ee
Eventually, equation \eqref{dpsi1_dr}, one of the equations \eqref{dp0_GEOS} to \eqref{dp0_PEOS} and equation \eqref{bc_r_0} are our boundary conditions.
\subsection{Computation method}\label{sec:method}
Applying the boundary conditions at the centre and the surface of the filament, one can solve equations \eqref{eq:rhoddr} to \eqref{eq:ddr} which demonstrate a disguised eigenvalue problem. This can be done by different methods. Here, we finite difference our equations over a 2000 point equally spaced mesh grid. This gives rise to a system of algebraic block-tridiagonal matrix equations which could be solved with any standard matrix solver. We take advantage of a flexible relaxation technique based on the Newton-Raphson-Kantorovich (NRK) standard algorithm \citep{Garaud-thesis}. This algorithm needs an initial guess to advance. For non-magnetic calculations, the algorithm converges rapidly after a few iterations using a reasonable guess for each of the dependent variables. For cases with the magnetic field, we start with the previous non-magnetic results as the initial guess. To calculate dispersion relation, we fix $\omega$ and consider the eigenvalue $k$ as a dependent variable. The algorithm successively adjusts $k$ along with the other dependent variables until it converges.
\section{Results}\label{sec:results}
\begin{table}
	\centering
	\begin{threeparttable}  
		\caption{Units of the filament radius $(R)$, the magnetic field strength $(B)$, the fastest perturbation growth rate $(\omega_{\rm fast})$ and the critical wave number $(k_{\rm critic})$ for a typical MC with the central density of 4$\times$10$^{-20}$ g cm$^{-3}$ and the thermal sound speed of 0.2 km s$^{-1}$.}
		\label{tab:units}
		\begin{tabular}{ccc}
			\hline\hline
			Parameter & Unit & Parameter = 1 is equivalent with\\ 
			\hline
			$R$ & $H$ & 0.035 pc\\ 
			$B$ & $c_{\rm s} \sqrt{4 \pi \rho_{\rm c}}$ & 14.180 $\upmu$G\\			
			$\omega_{\rm fast}$ & $\sqrt{4 \pi G \rho_{\rm c}}$ & $5.780$  $\text{Myr}^{-1}$\\
			$k_{\rm critic}$ & $1/H$ & 28.409 $\text{pc}^{-1}$\\			
			\hline
		\end{tabular}
	\end{threeparttable}
\end{table}

\begin{figure*}
	\centering
	\includegraphics[scale=0.46]{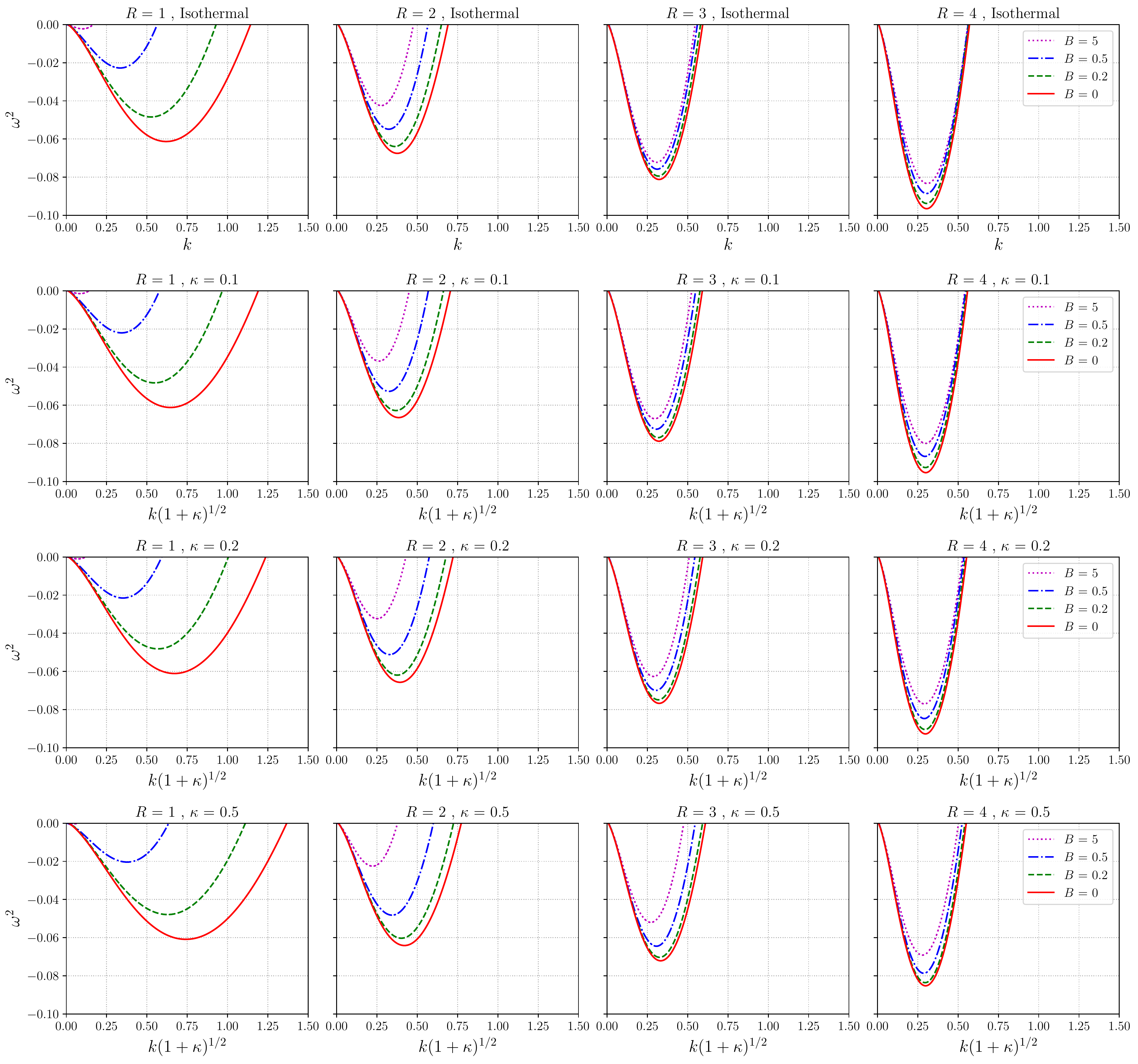}
	\caption{The GEOS dispersion relations for $\kappa = 0$ (isothermal), $ 0.1, 0.2, 0.5$, different filament radii ($R$). $R=1$ is equivalent with $R\simeq0.035$ pc and $B=5$ is equivalent with $B\simeq71$ $\upmu$G. In each panel the vertical axis is $\omega^2$ in the unit of $4 \pi G \rhoc$ and the horizontal axes are the wave numbers ($k$) which is multiplied by $(1+\kappa)^{1/2}$ to account for using the thermal sound speed as the velocity unit. Units are given in Table \ref{tab:units}. For $R = 1$, $B=5$ and $\kappa \geqslant 0.2$, $|\omega^2|$ is too small to be recognized.}
	\label{fig:Gehman_R_selected_Plot_All_kappa}
\end{figure*}
By considering the effect of external pressure of the environment which confines the filament boundary, we try to study the stability of the filament under influence of various magnetic field strengths and different EOSs. For each EOS dispersion relation, we select the minimum of $\omega^2$ or equivalently the maximum of $|\omega^2|$ and substitute it in $\omega_{\rm fast} = \sqrt{|\omega^2|}$ to find the fastest perturbation growth rate. A system with larger $\omega_{\rm fast}$ is more prone to the instability and vice versa. So $\omega_{\rm fast}$ is a good indicator in order to study the instability of filaments. Also we define $k_{\rm critic}$ as the non-zero wave number corresponding to $\omega=0$, here which is also the largest unstable one. The wave numbers with $\omega^2<0$ and $k<k_{\rm critic}$ are unstable. The unit of the filament radius $(R)$, the magnetic field strength $(B)$, the fastest perturbation growth rate $(\omega_{\rm fast})$ and the critical wave number $(k_{\rm critic})$ and their equivalent values when they are equal to 1 are summarized in Table \ref{tab:units}.
We adopt different values for the filament radius, the magnetic field strengths, the turbulence parameters ($\kappa$ in the GEOS) and the polytropic indices ($n$ in the PEOS) to get as high as possible resolution for dispersion relations in the different EOSs. For example, we set $R = 0.25, 0.5, 0.75, 1, 1.25, ..., 3.75, 4, 4.25$, $B = 0, 0.01, 0.02, 0.05, 0.1, 0.2, 0.5, 1, 2, 5, 10$ and $\kappa = 0, 0.1, 0.2, 0.5, 1$ for the GEOS to provide the dispersion relations and investigate the effect of $R$, $B$ and $\kappa$ on the instability of this type of filaments. Albeit settings $\kappa$ to the values larger than 1 leads the minimum value of $|\omega^2|$ to become about zero for the radii less than 1 and for all the magnetic field strengths. It is worth noting that $R = 1$ is equivalent with $R =H \simeq 0.035$ pc and considering the density of 4$\times$10$^{-20}$ g cm$^{-3}$ and the thermal sound speed of 0.2 km s$^{-1}$, $B = 1$ is equivalent with $B \simeq 14.2$ $\upmu$G. Similarly we set $R = 0.25, 0.5, 0.75, 1, 2, 5, 10, 20, 50$ and $A = 0.2$ for the MPEOS with the same magnetic fields to achieve dispersion relations. For the PEOS filaments, we use the same $R$ and $B$ values as for the MPEOS and $n = -1.5, -2, -3, -4$.

\begin{figure*}
	\centering
	\includegraphics[scale=0.37]{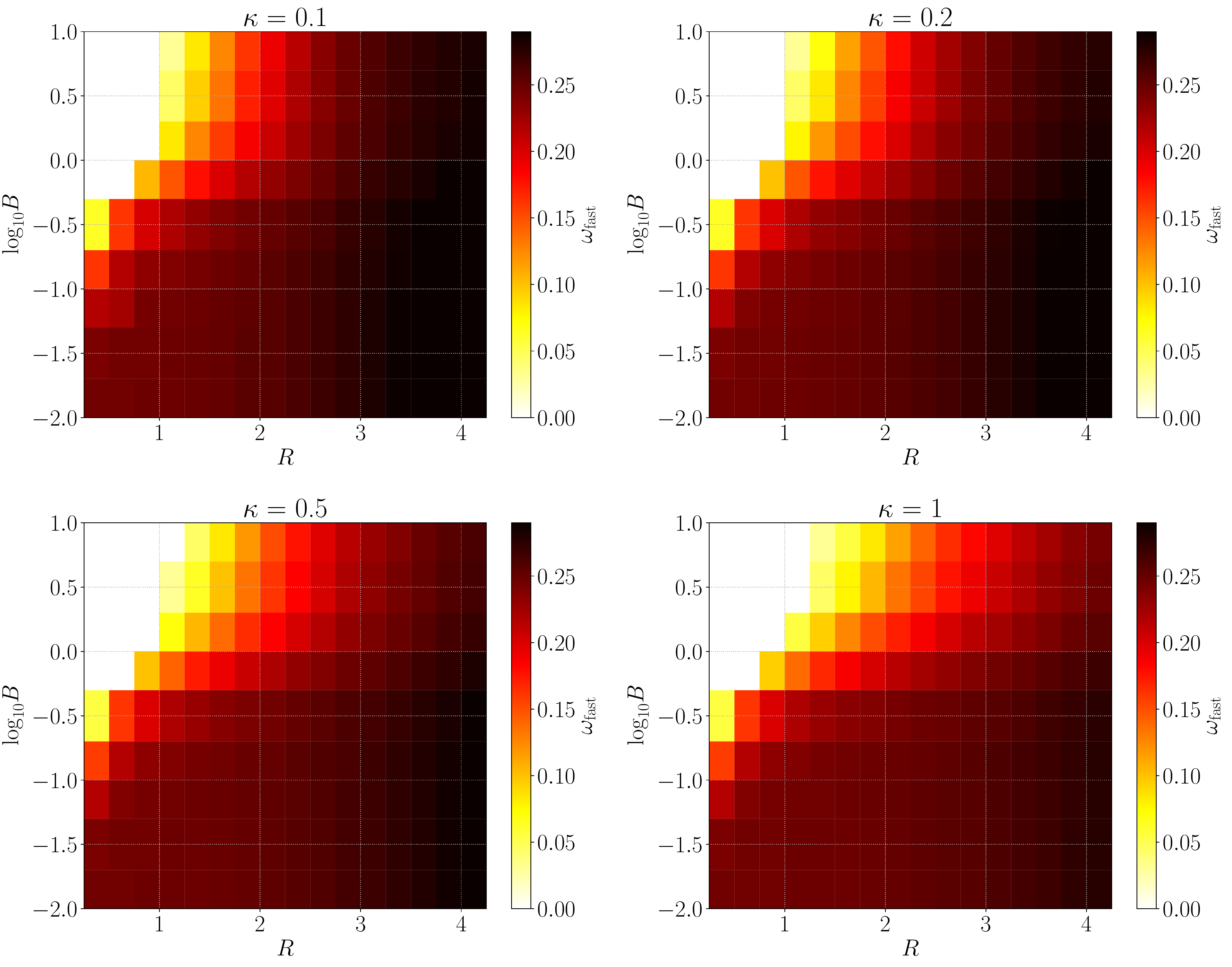}
	\caption{Effect of the filament radius ($R$) and the magnetic field ($B$) on the GEOS filament instability with $\kappa=0.1$ (upper left panel), $\kappa=0.2$ (upper right panel), $\kappa=0.5$ (lower left panel) and $\kappa=1$ (lower right panel). In each panel, the horizontal axis shows the radius of filaments, the vertical axis shows the logarithm of magnetic field strength and the colour bar represents $\omega_{\rm fast}$. Units are given in Table \ref{tab:units}. The darker shaded areas are more unstable.}
	\label{fig:Gehman_meshgrid_All_kappa_omega_fast}
\end{figure*}
Fig. \ref{fig:Gehman_R_selected_Plot_All_kappa} shows the dispersion relations for the IEOS (a GEOS with $\kappa=0$) and the GEOS. Although we calculate the dispersion relations using so many values for the filament radii, the magnetic field strengths and the turbulence parameters $(\kappa)$, here, we draw the dispersion relations for only $R = 1, 2, 3, 4$ and $B = 0, 0.2, 0.5, 5$ in order to observe their impact on the stability of the filament. Panels in each column of Fig. \ref{fig:Gehman_R_selected_Plot_All_kappa} show specific radii and ones in each row represent filaments with a specific turbulence parameter. As \cite{Nagasawa87}, \cite{Gehman2} and \citetalias{Rad2018} reported, because of the magnetic field disability to prevent fluid contraction along the field direction, dispersion relation curves are nearly overlapped for the magnetic fields larger than $B \simeq 5$ or $\simeq 71$ $\upmu$G. We compare the dispersion relations of our models with the work by \cite{Nagasawa87} and find that the two studies reproduce fairly similar results.
\begin{figure*}
	\centering
	\includegraphics[scale=0.37]{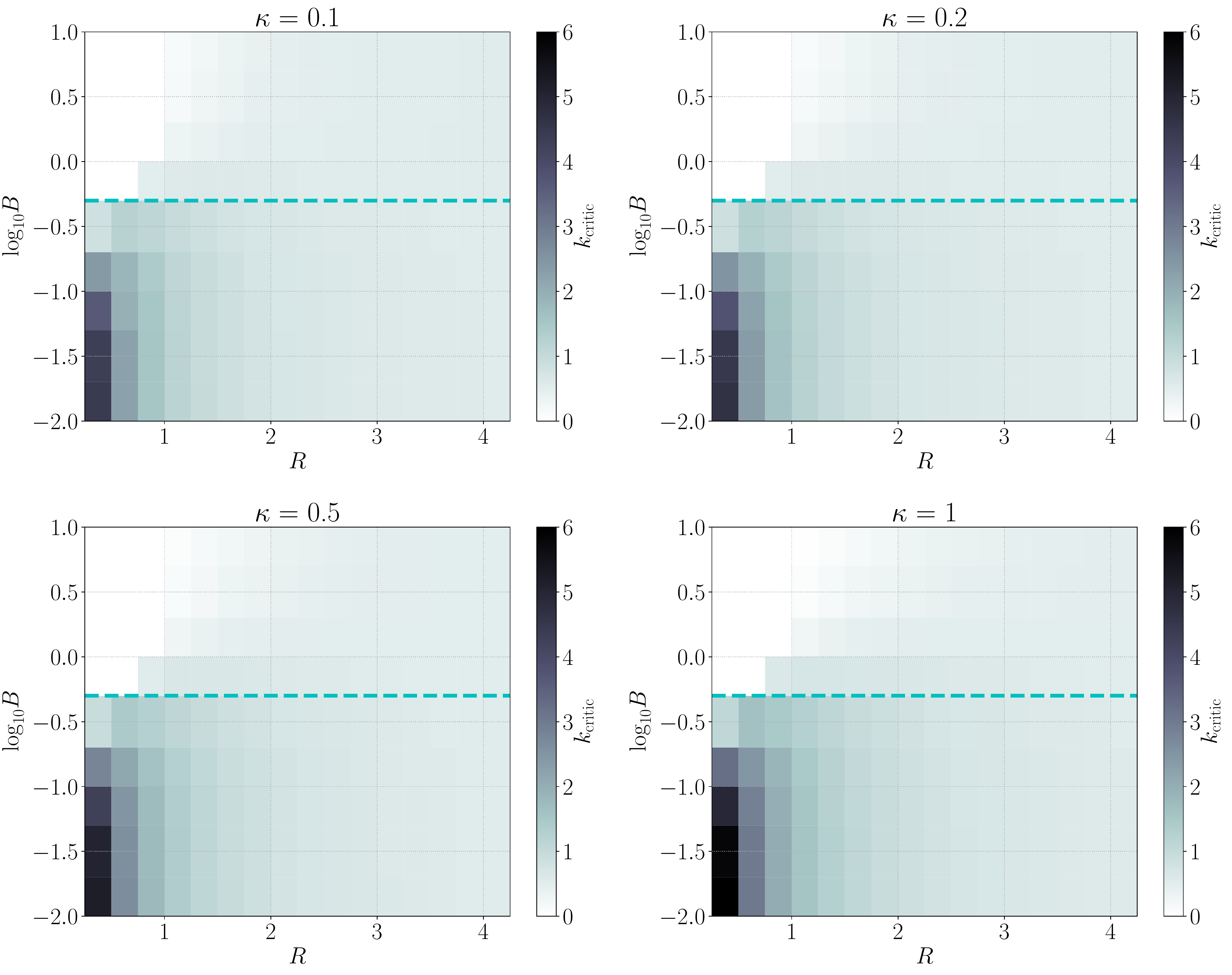}
	\caption{Effect of the filament radius ($R$) and the magnetic field ($B$) on the critical wave number ($k_{\rm critic}$) in the GEOS dispersion relations with $\kappa=0.1$ (upper left panel), $\kappa=0.2$ (upper right panel), $\kappa=0.5$ (lower left panel) and $\kappa=1$ (lower right panel). The horizontal and vertical axis in each panel shows the radius of filaments and the logarithm of magnetic field strength, respectively and the colour bar represents $k_{\rm critic}$. Units are given in Table \ref{tab:units}. The horizontal dashed line separates models with the magnetic fields larger than $B \simeq 0.5$ or $\log_{10}B \simeq -0.3$ from others in each panel.}
	\label{fig:Gehman_meshgrid_All_kappa_k_critic}
\end{figure*}
In all the panels of Fig. \ref{fig:Gehman_R_selected_Plot_All_kappa} it is obvious that increasing the magnetic field leads to more stability by decreasing $\omega_{\rm fast}$. This also causes to the reduction of $k_{\rm critic}$ value. We can see for each $\kappa$ and in a specific magnetic field, when the radius of a filament increases, $\omega_{\rm fast}$ increases as well and so the filament becomes more unstable. With a focus on the radius of filaments, the instability of filaments with smaller radii are more sensitive to the magnetic field strength than the larger ones. On the other hand for a filament with a specific radius and in a fixed magnetic field, when $\kappa$ increases, $\omega_{\rm fast}$  decreases and so the filament becomes more stable. Interestingly, this is in contrast to the behaviour of infinite filaments (here by infinite we mean large in radius) (\citetalias{Rad2018}). In addition, clearly none of the models represent the radial instability (RI) which occurs at a non-zero $\omega^2$ corresponding to $k=0$, in any panel that is again in contrast to the infinite filaments (\citetalias{Rad2018}). 

To understand the problem with more details, we performed very large surveys of stability analysis that include the effect of various magnetic field strengths, different radii and several values of $\kappa$. Fig. \ref{fig:Gehman_meshgrid_All_kappa_omega_fast} represents the result of four surveys for the GEOSs with $\kappa=0.1, 0.2, 0.5$ and 1. In each panel, the colour bar shows $\omega_{\rm fast}$. The darker colour the larger $\omega_{\rm fast}$ and the more unstable filament. Areas with lighter colour, have lower $\omega_{\rm fast}$ and so represent more stable models. It is clear that the lower right region of each panel which represents the larger radii i.e. the thicker filaments in the lower magnetic field strength, is the most unstable area. In contrast, the upper left region in each panel which indicates the thinner filaments in the stronger magnetic fields are the most stable ones. In particular, if we compare panels with each other, we find that when $\kappa$ increases, the colour of a specific area becomes lighter that means the corresponding filament is more stable. It is also of the note that there are some areas with different radii and magnetic fields but with the same colour i.e. the same $\omega_{\rm fast}$. This means that just by knowing the $\omega_{\rm fast}$, it is not possible to find the magnetic field in a filament or its radius exclusively.

In Fig. \ref{fig:Gehman_meshgrid_All_kappa_k_critic} we can see the effect of filament radius and the magnetic field on $k_{\rm critic}$ in the GEOS dispersion relation for the same $\kappa$ values as the Fig. \ref{fig:Gehman_meshgrid_All_kappa_omega_fast}. This figure shows that $k_{\rm critic}$ has very weak dependency on $\kappa$ on the whole. Furthermore, if we compare this figure with Fig. \ref{fig:Gehman_meshgrid_All_kappa_omega_fast}, we find that models that are in the upper left corner of the panels i.e. the thinner filaments in the stronger magnetic fields, which have the smallest $\omega_{\rm fast}$, have also the smallest $k_{\rm critic}$. On the other hand, there is an area with larger $k_{\rm critic}$ that contains the thinner filaments in the weaker magnetic field strengths (the lower left region in all the panels of Fig. \ref{fig:Gehman_meshgrid_All_kappa_k_critic}). It is also observed that in all the panels, by increasing the filament radii in magnetic fields $B \gtrsim 0.5$ ($\log_{10} B \gtrsim -0.3$, indicated by horizontal dashed line), $k_{\rm critic}$ also increases. However in $B < 0.5$, an inverse trend is observed. $B = 0.5$ is equivalent with $B \simeq 7.1$ $\upmu$G. Notwithstanding this different treatment in the stronger and the weaker magnetic field regimes, $k_{\rm critic}$ is converged to about 0.5 for $R\gtrsim3.5$ in all the panels.

\begin{figure*}
	\centering
	\includegraphics[scale=0.41]{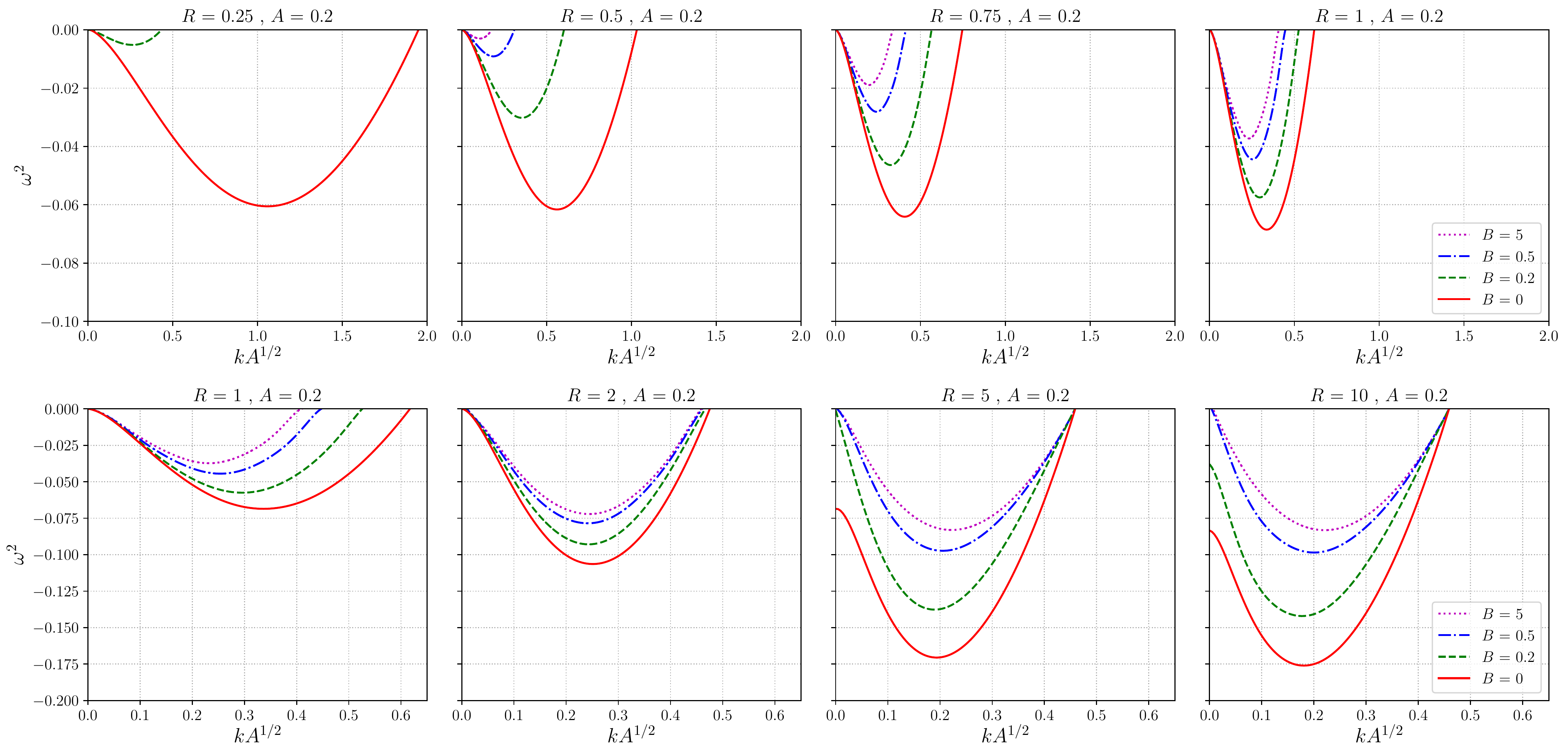}
	\caption{The MPEOS dispersion relations for different filament radii ($R$) and magnetic field strengths ($B$). The vertical axes are as Fig. \ref{fig:Gehman_R_selected_Plot_All_kappa}. The horizontal axes are the wave numbers $(k)$ multiplied by $A^{1/2}$ to account for using the thermal sound speed as the velocity unit. Units are given in Table \ref{tab:units}. For $R = 0.25$ and $B > 0.2$, $|\omega^2|$ is too small to be recognized. Note to the different scale of the horizontal axes in the first and the second rows. The $R=1$ panel is repeated in the second row in order to better compare with the larger radii.}
	\label{fig:MP_R_selected_Plot}
\end{figure*}
\begin{figure*}
	\centering
	\includegraphics[scale=0.37]{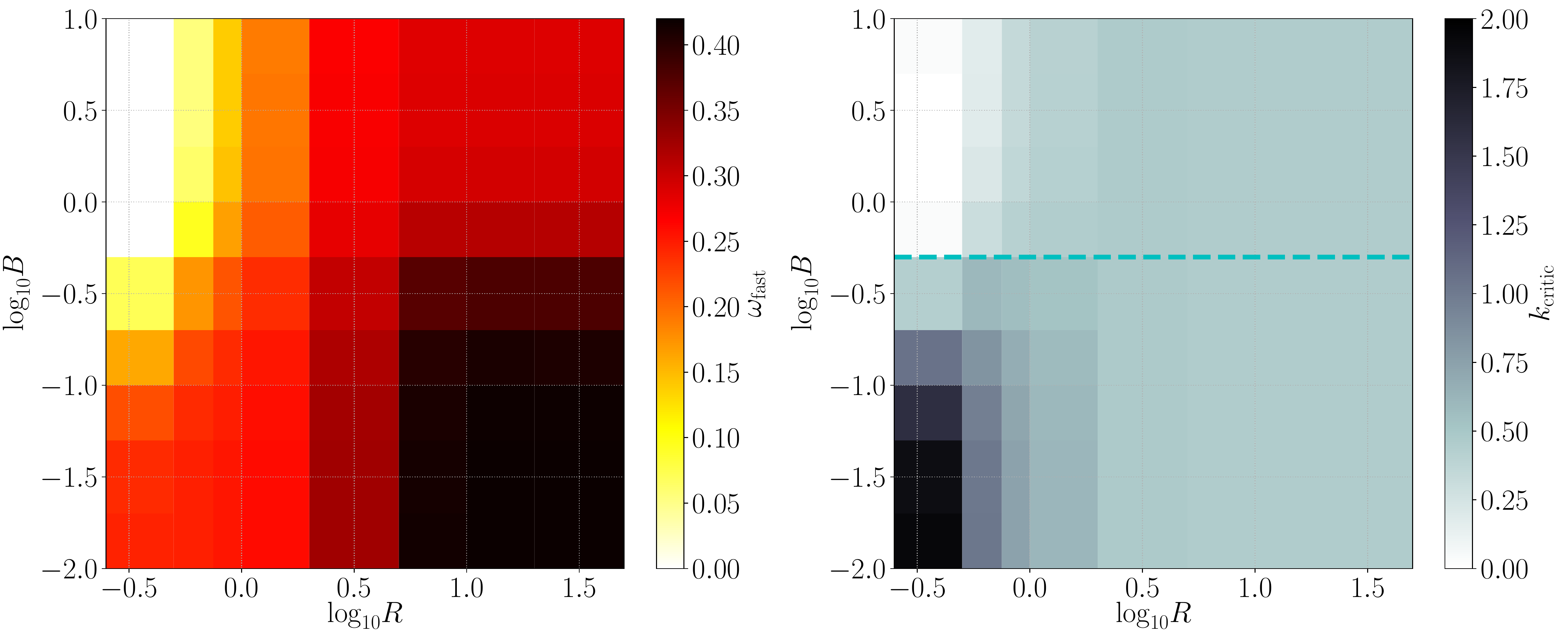}
	\caption{Left-hand panel: Effect of the filament radius ($R$) and the magnetic field ($B$) on the instability of the MPEOS filaments. In each panel the horizontal axis shows the logarithm of filament radius, the vertical axis shows the logarithm of magnetic field strength and the colour bar represents $\omega_{\rm fast}$. The units of $R$, $B$ and $\omega_{\rm fast}$ are as Fig. \ref{fig:Gehman_meshgrid_All_kappa_omega_fast}. The darker shaded areas are more unstable. Right-hand panel: Effect of the filament radius ($R$) and the magnetic field ($B$) on the critical wave number ($k_{\rm critic}$) in the MPEOS dispersion relation. The horizontal and vertical axes are as the left-hand panel and the unit of $k_{\rm critic}$ are as Fig. \ref{fig:Gehman_meshgrid_All_kappa_k_critic}. The horizontal dashed line separates models with the magnetic fields larger than $B \simeq 0.5$ or $\log_{10}B \simeq -0.3$ from others in each panel.}
	\label{fig:MP_meshgrid_All}
\end{figure*}
\begin{figure*}
	\centering
	\includegraphics[scale=0.46]{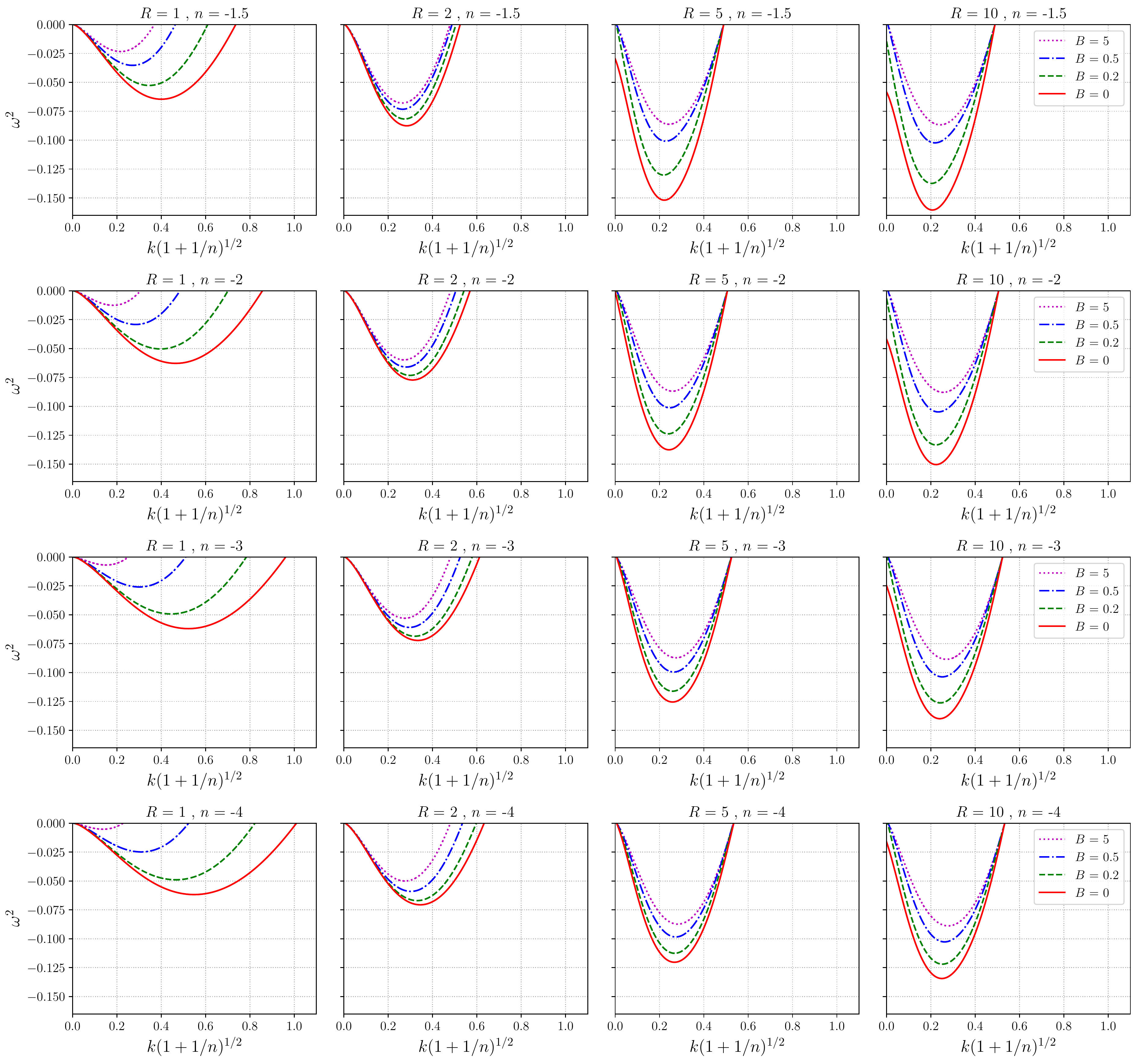}
	\caption{The PEOS dispersion relations for polytropic indices ($n$) of $-1.5, -2, -3$ and $-4$, different filament radii ($R$) and magnetic field strengths ($B$). The vertical axes are as Fig. \ref{fig:Gehman_R_selected_Plot_All_kappa}. The horizontal axes are the wave numbers $(k)$ multiplied by $(1+1/n)^{1/2}$ to account for using the thermal sound speed as the velocity unit. Units are given in Table \ref{tab:units}.}
	\label{fig:poly_R_selected_Plot_All_n}
\end{figure*}
\begin{figure*}
	\centering
	\includegraphics[scale=0.37]{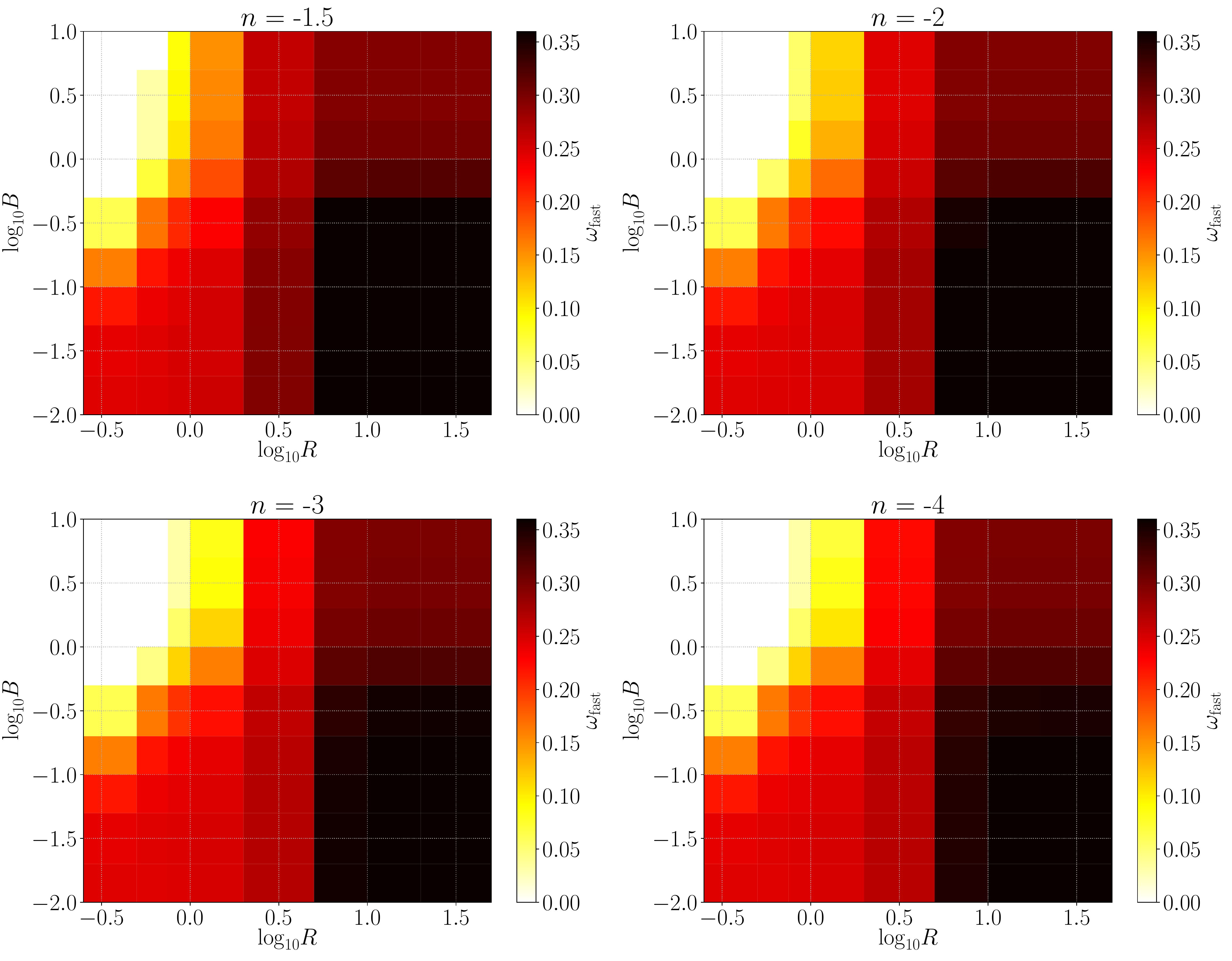}
	\caption{Effect of the filament radius ($R$) and the magnetic field ($B$) on the PEOS filaments instability with $n=-1.5$ (upper left panel), $n=-2$ (upper right panel), $n=-3$ (lower left panel) and $n=-4$ (lower right panel). In each panel the horizontal axis shows the logarithm of the filament radius, the vertical axis shows the logarithm of magnetic field strength and the colour bar represents $\omega_{\rm fast}$. The units of $R$, $B$ and $\omega_{\rm fast}$ are as Fig. \ref{fig:Gehman_meshgrid_All_kappa_omega_fast}. The darker shaded areas are more unstable.}
	\label{fig:poly_meshgrid_All_n_omega_fast}
\end{figure*}
Fig. \ref{fig:MP_R_selected_Plot}, demonstrates the dispersion relations for the MPEOS. Similar to the GEOS, the dispersion relations were plotted for several values of the filament radius, the same magnetic field strengths and $A=0.2$. Because for $R\gtrsim10$, there are no noticeable changes in the dispersion relations, in this figure, we select only $R=0.25, 0.5, 0.75, 1, 2, 5, 10$ in order to see $\omega_{\rm fast}$ and $k_{\rm critic}$ variations clearly. Each panel shows the dispersion relation for the aforementioned filament radii. Like the GEOS dispersion relations (see Fig. \ref{fig:Gehman_R_selected_Plot_All_kappa}), for all the radii, increasing the magnetic field, increases the stability of the filament. Yet also the stabilization effect due to the magnetic field is saturated for $B > 5$ \citep[\citetalias{Rad2018}]{Nagasawa87, Gehman2}. In small radii, the effect of magnetic field on the dispersion relation is stronger. As an example a filament with the radius of $R=0.5$ is almost stable when $B=5$. This is also the case for a smaller radius of $R=0.25$, but for an order of magnitude weaker magnetic field. This filament is completely stable when $B=5$. It should be also noted that for the radius of $R=5$ in the absence of magnetic field and for the radius of $R=10$ also in presence of a relatively weak magnetic field of $B=0.2$ the filament is radially unstable.
In addition, it is noticeable that RI disappears entirely when $B\geqslant 0.5$.
\begin{figure*}
	\centering
	\includegraphics[scale=0.37]{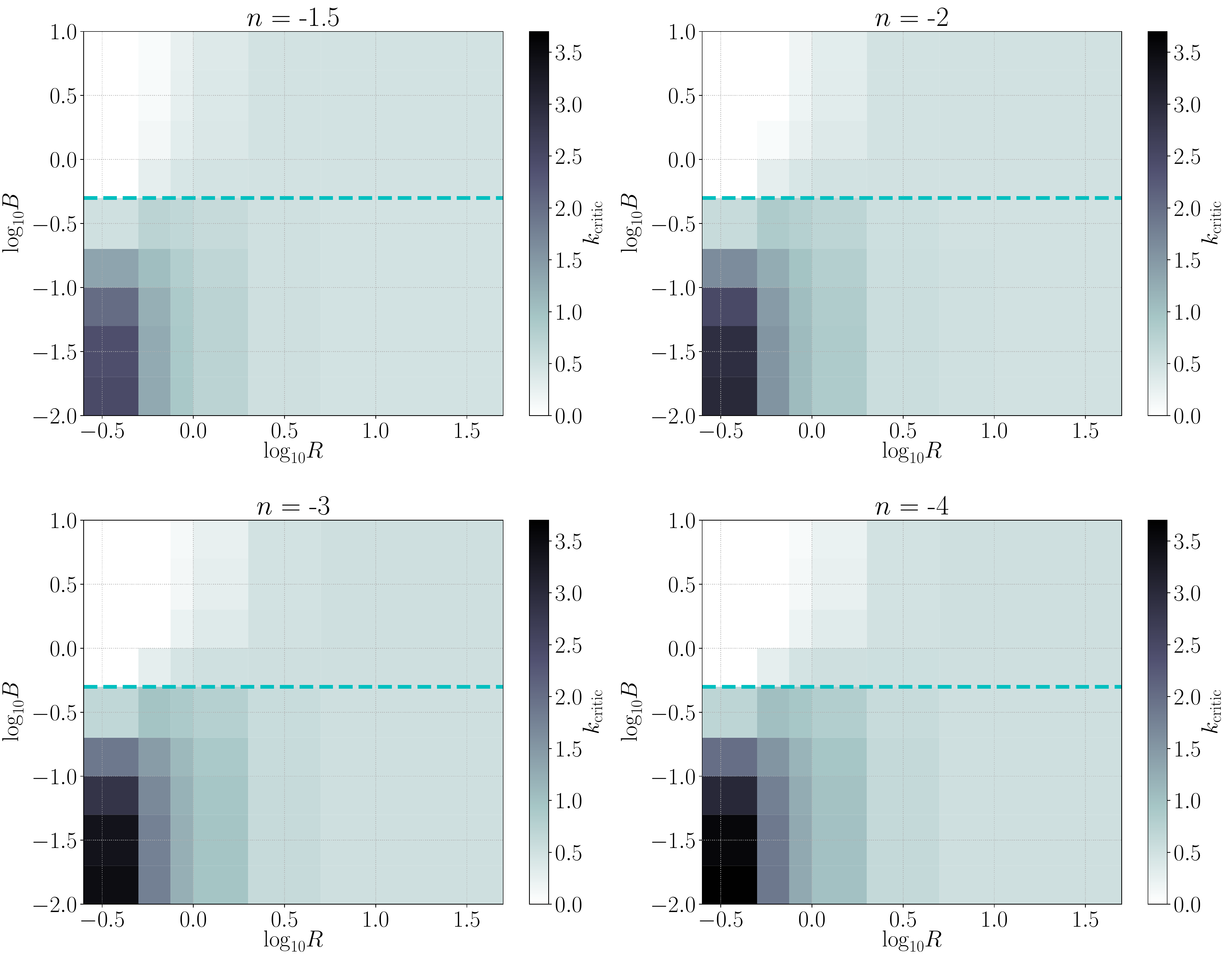}
	\caption{Effect of the filament radius ($R$) and the magnetic field ($B$) on the critical wave number ($k_{\rm critic}$) in the PEOS dispersion relations with $n=-1.5$ (upper left panel), $n=-2$ (upper right panel), $n=-3$ (lower left panel) and $n=-4$ (lower right panel). In each panel the horizontal axis shows the logarithm of filament radius, the vertical axis shows the logarithm of magnetic field strength and the colour bar represents $k_{\rm critic}$. The units of $R$, $B$ and $k_{\rm critic}$ are as Fig. \ref{fig:Gehman_meshgrid_All_kappa_k_critic}. The horizontal dashed line in each panel separates the magnetic fields larger than or smaller than $B \simeq 0.5$ or $\log_{10}B \simeq -0.3$.}
	\label{fig:poly_meshgrid_All_n_k_critic}
\end{figure*}

Much the same as the GEOS, the stability analysis is elaborated in the left-hand panel of Fig. \ref{fig:MP_meshgrid_All} which shows instability of the MPEOS filaments with different radii in various magnetic fields with more details. Comparing this panel with those of Fig. \ref{fig:Gehman_meshgrid_All_kappa_omega_fast}, one can see that the general treatment is the same but the MPEOS is a little more unstable than the GEOSs. By incrementally increasing the magnetic field, same as the GEOS, all the filaments become more stable gradually. Moreover, the upper left region has the lightest colour and manifestly shows that the most stable filaments are the thinnest ones in the strongest magnetic field regime. Furthermore, it should be noted that similar to the Fig. \ref{fig:Gehman_meshgrid_All_kappa_omega_fast} for the GEOS, some models have the same $\omega_{\rm fast}$ in spite of having different radii because of the effect of magnetic field. As it could be observed in the panel, like the GEOS, increasing the radius leads to more instability regardless of the magnetic field strength. For $R>5$, increasing the radius has not noticeable effect on the instability for all the magnetic field strength. The right-hand panel of Fig. \ref{fig:MP_meshgrid_All} shows the effect of filament radius and the magnetic field on  $k_{\rm critic}$. Here, the overall picture is very similar to the GEOSs, specially one with $\kappa = 0.1$. The only difference is that the very thin MPEOS filaments in the absence or presence of very weak magnetic fields, have $k_{\rm critic}$ almost twice larger than their GEOS counterparts.

Fig. \ref{fig:poly_R_selected_Plot_All_n} represents the dispersion relations for the PEOS. As already mentioned in $\S$ \ref{sec:PEOS}, we set $n=-1.5, -2$ and $-3$ as the polytropic indices and use $R$ and $B$ values same as in the MPEOS. We also compute the dispersion relation for $n=-4$ in order to examine the effect of smaller $n$ on the instability. In this figure, panels in each row indicate filaments with a specific polytropic index, while those in each column show specific radii. The filaments radii $R>10$, do not show noticeable changes in their dispersion relations, so we select only $R=0.25, 0.5, 0.75, 1, 2, 5$ and $10$ in order to see their effect on the filaments instability. Nevertheless, to see the impact of these parameters on the shape of dispersion relations more clearly, we draw the plots only for $R = 1, 2, 5$ and $10$ and $B = 0, 0.2, 0.5$ and $5$. Here, once more the magnetic field strength $B>5$ causes the dispersion relation to nearly overlap similar to the GEOS and the MPEOS \citep[\citetalias{Rad2018}]{Nagasawa87, Gehman2}. In all the panels of Fig. \ref{fig:poly_R_selected_Plot_All_n} it is obvious that increasing the magnetic field leads to the more stability by decreasing $\omega_{\rm fast}$ and for $R\lesssim5$, reduction of $k_{\rm critic}$ value. {\citetalias{Rad2018}} showed that the infinite filaments with the PEOS are prone to the radial instability and a strongly enough magnetic field could suppress the instability. Focusing on the $R=5$ and $R=10$ panels, we see that this is also the case for the pressure confined filaments with PEOSs and relatively large radii. It should be noted that the smaller radius is radially unstable just for $n=-1.5$ while the larger one is unstable for all the polytropic indices. The occurrence of RI could be probably due to this fact that the effective sound speed decreases when the density of the filament increases. For $R<5$ there is no sign of RI in the dispersion relations.

In the following, the effect of $R$, $B$ and $n$ on the instability of filaments with the PEOS is studied by leveraging the result of four large surveys in more details. Fig. \ref{fig:poly_meshgrid_All_n_omega_fast} illustrates the outcome for $\omega_{\rm fast}$. The upper left stable region is a common clear feature of all the panels which states that similar to the previous GEOS and MPEOS, the thinner filaments in the stronger magnetic fields are the most stable ones. Looking at the panels, it is clear that the stability patterns are more or less similar to the GEOS and MPEOS. All the four panels have a distinctive more stable region at the upper left and an unstable region at the lower right corner. By decreasing $n$, the former becomes a little larger while the latter fades out. This could indicate that the softer PEOSs (ones with larger $n$) are more unstable, possibly because these filaments have more mass per unit length. Moreover, like the GEOS and MPEOS by increasing the radius of a filament in a constant magnetic field, the stability decreases until $R=10$ or $\simeq 0.35$ pc where after this radius, the stability does not change noticeably. Regarding the stabilizing effect of magnetic field, the stability of filaments with smaller radii are more sensitive to the magnetic field strength than the larger ones. It is also worth noting that similar to Fig. \ref{fig:Gehman_meshgrid_All_kappa_omega_fast} and \ref{fig:MP_meshgrid_All}, there is a degeneracy in determining $B$ and $R$ from a specific $\omega_{\rm fast}$.

Fig. \ref{fig:poly_meshgrid_All_n_k_critic} exhibits how the filament radius and the magnetic field could affect $k_{\rm critic}$ in the PEOSs with different $n$. The results are very similar to the GEOS and MPEOS, however, one can see that for the thinner filaments in the low magnetic field regime, the GEOS has the greatest $k_{\rm critic}$ while the MPEOS has the smallest one. In addition, by decreasing $n$, one can see that for the thinner filaments in the low magnetic field regime, $k_{\rm critic}$ is a little increases. For all the models, $k_{\rm critic}$ is converged to $\simeq 0.5$ after $R>5$.

To further study the relationship between the critical wavelength of the fragmentation $(\lambda_{\rm critic}=2 \pi / k_{\rm critic})$ and the EOS, we calculate $\lambda_{\rm critic}$ for the selected $R$, $B$, $\kappa$, $n$ and $A$ (Table \ref{tab:result}). In this table because $k_{\rm critic}$ is about 0 for some radii, magnetic fields and EOSs, we can not calculate $\lambda_{\rm critic}$ for those models (indicated as N/A in the table). As expected, $\lambda_{\rm critic}$ is more sensitive to the magnetic field and the type of EOS for thinner filaments. Also difference between the maximum and the minimum values of $\lambda_{\rm critic}$ in the various EOSs is greater for the filaments with smaller radii.
	
\begin{table}
	\centering
	\begin{threeparttable}  
	\caption{$\lambda_{\rm critic}$ values in parsecs for the sample filaments radii ($R$), the magnetic field strengths ($B$), the turbulence parameters ($\kappa$) in the GEOS, the polytropic indices ($n$) in the PEOS and the empirical constant ($A$) in the MPEOS.}
	\label{tab:result}
	\begin{tabular}{ccccccc}
		\hline\hline
		$ $ & $ $ &\multicolumn{2}{c}{$\kappa$} &\multicolumn{2}{c}{$n$} &\multicolumn{1}{c}{$A$}\\ 
		\cmidrule(lr){3-4} \cmidrule(lr){5-6} \cmidrule(lr){7-7}
		$R$ (pc) & $B$ ($\upmu$G) & 0.0 & 1.0 & -1.5 & -4.0 & 0.2 \\ 
		\hline
		0.018&00.0&0.102&0.073&0.171&0.117&0.214\\
		0.018&07.1& N/A\tnote{a} & N/A &0.771&0.801&0.713\\
		0.018&70.9& N/A & N/A &2.697& N/A &1.250\\
		\vspace*{-2.5mm}			
		&&&&&&\\
		0.035&00.0&0.193&0.142&0.301&0.219&0.358\\
		0.035&07.1&0.392&0.314&0.478&0.422&0.491\\
		0.035&70.9&1.324&6.505&0.599&0.970&0.536\\
		\vspace*{-2.5mm}			
		&&&&&&\\
		0.100&00.0&0.369&0.330&0.446&0.394&0.480\\
		0.100&07.1&0.395&0.381&0.452&0.416&0.480\\
		0.100&70.9&0.408&0.520&0.453&0.425&0.480\\
		\vspace*{-2.5mm}			
		&&&&&&\\
		0.175&00.0&0.391&0.419&0.452&0.414&0.482\\
		0.175&07.1&0.392&0.433&0.452&0.415&0.482\\
		0.175&70.9&0.393&0.451&0.452&0.416&0.482\\
		\hline
	\end{tabular}
	\begin{tablenotes}  
		\item[a] Not available data for this model (see the text).
	\end{tablenotes}  
	\end{threeparttable}
\end{table}

\section{Summary and conclusion}\label{sec:conclusion}
Filamentary structures seem to be a natural early stage in formation of stars and clusters of stars. This has stimulated many investigations regarding the properties and evolution of these structures. In a pioneering work, \cite{Nagasawa87} showed that the pressure-confined filaments are gravitationally unstable for a specific range of wavelengths and a poloidal magnetic field can increase their stability and interestingly entirely stabilize them if the filaments are thin enough.

Recent observations show that the IEOS is not always the best EOS for interpreting the filaments properties. Building on the work by \citet{Nagasawa87}, \citetalias{Rad2018} studied the instability of filamentary MCs without the effect of external pressure, with the previously proposed non-isothermal EOSs, namely the GEOS, the MPEOS and the PEOS.

In this paper, in a continuation of the previous work by \citetalias{Rad2018}, we have added the effect of external pressure. To this aim, we use these three non-isothermal EOSs (described in $\S$ \ref{sec:EOS}) in order to study the instability of magnetized pressure-confined filaments.
We solve the equations as mentioned earlier in $\S$ \ref{sec:method} and extract the dispersion relations for these non-isothermal filaments with various radii ($R$) and magnetic field ($B$). Moreover, by exploiting the growth rate of the fastest growing mode ($\omega_{\rm fast}$) as a gravitational instability indicator, we are able to investigate the effect of filament radius, magnetic field and type of EOS on the instability of the filaments. In summary, the results show that:
\begin{enumerate}
	\item Similar to the infinite filaments, for all the EOSs, increasing the magnetic field strength, makes the pressure-confined filaments more stable.
	\item The instability in the thinner filaments is  more sensitive to the magnetic field strength than the thicker ones.
	\item Unlike the infinite filaments, for the GEOS pressure-confined models considered in this study which have $R\lesssim 0.15$ pc, for all the radii (specially larger ones), in a fixed magnetic field, when $\kappa$ increases, the filaments become more stable.
	\item For all the EOSs, the thinner filaments are totally stabilized in an even intermediate magnetic field strength (e.g. models with $R\lesssim 0.03$ pc in $B\gtrsim 14$ $\upmu$G), while for the thicker ones this effect is suppressed for the magnetic field stronger than $B\simeq 70$ $\upmu$G.
	\item There is no RI in the GEOS pressure-confined filaments. This is in contrast to the infinite GEOS filaments. 
	\item In the absence of magnetic field, the MPEOS and the PEOS with $n=-1.5$ and $R\gtrsim 0.17$ pc are radially unstable.
	The twice broader filaments of these two EOSs can also be radially unstable in presence of a weak magnetic field $B\simeq 3\upmu$G. The RI in the other less softer PEOSs ($n=-2, -3$ and $-4$) with the radius $R=0.35$ pc is still suppressed by a weak magnetic fields of $B\simeq7 \upmu$G for the first and $B\simeq 3\upmu$G for the next two ones.
	\item In the PEOS, decreasing $n$ has the same effect on the filament instability as the increasing $\kappa$ in the GEOS. 
	\item Comparing the filaments with the same radius and in the same magnetic field, the MPEOS filaments are the most unstable ones, because of their softer EOS.
	\item The minimum spacing distance between clumps in filamentary MCs often is compared with $\lambda_{\rm critic}$ \citep[e.g.][]{2011A&A...533A..34H,2016MNRAS.456.2041C,2020arXiv201207738Z} and demonstrates diverse ranges of length (see Table \ref{tab:result}). The predicted $\lambda_{\rm critic}$ is clearly dependent on the filament radii, the EOS and the magnetic field strength. This dependency is more pronounced for the thinner filaments and is completely strong for the thinnest ones. Caution is needed in interpreting this length scale. More specifically, it is interesting to investigate the fragmentation space within the thinner filaments \citep[e.g.][]{2014A&A...569A..11S}.
	\item In all the models it is observed that by decreasing the filament radius (which means the higher external pressure), $\omega_{\rm fast}$ decreases or equivalently the minimum time needed for the fragmentation ($\tau_{\rm min}=1 / \omega_{\rm fast}$) increases. This could correspond to a longer time needed for a clump to become unstable and finally form protostars . Interestingly, this has been also reported by \citet{anathpindika2020filament} recently. They have performed hydrodynamical simulation of accreting filaments in a medium with different external pressure and have shown that a higher external pressure leads to a lower star formation rate.
	\item By comparing $\omega_{\rm fast}$ for the PEOSs, one can see that the softer ones have smaller $\tau_{\rm min}$. Remarkably, in agreement with this result, in a hydrodynamical simulation of an initially uniform polytropic gas within a periodic box and driving turbulence, \citet{Federrath_2015} derived that the star formation rate increases for the softer PEOSs.
\end{enumerate}

\section*{Data availability}
No new data were generated or analysed in support of this research.

\section*{Acknowledgements}
Mohammad Mahdi Motiei and Mohammad Hosseinirad thank Mahmood Roshan and Najme Mohammad-Salehi for useful discussions. The authors thank the anonymous referees for the careful reading of the manuscript and their insightful and constructive comments. This research made use of \texttt{Scipy}~\citep{scipy}, \texttt{Jupyter}~\citep{jupyter} and \texttt{Numpy}~\citep{numpy}. All figures were generated using \texttt{Matplotlib}~\citep{matplotlib}. Also we have made extensive use of the NASA Astrophysical Data System Abstract Service. This work was supported by the Ferdowsi University of Mashhad under grant no. 50729 (1398/06/26).

\bibliography{my_paper} 
\appendix
\section{Dimensionless form of the equations of state}\label{sec:dimensionless}
We convert quantities and equations to the dimensionless ones as \citetalias{Rad2018}. The units are
\be [\rho]=\rho_{\rm c},\ee
\be [t]={\sqrt{4\pi G[\rho]}}^{-1},\ee
\be [p]=p_c,\ee
\be [\bvec{u}]=\sqrt{\dfrac{[p]}{[\rho]}},\ee
\be [r]=[t][\bvec{u}],\ee
\be [\psi]=[\bvec{u}]^2.\ee
The velocity unit is equal to the isothermal sound speed $\cs$ for the IEOS and the GEOS. For the MPEOS and the PEOS, it is assumed to be $\cs$. The magnetic field unit is defined as 
\be [\bvec{B}]=\sqrt{4\pi[p]},\ee
$B=1$ is equivalent with $B \simeq 14.2$ $\upmu$G.
Using these units, we can rewrite the analytical solution of the density and gravitational potential of the isothermal filament as
\begin{equation}
\rho(r)=\bigg(1+\dfrac{r^2}{8}\bigg)^{-2}
\label{eq:rho_nond},
\end{equation}
and
\begin{equation}
\psi(r)=2\ln\bigg(1+\dfrac{r^2}{8}\bigg).
\label{eq:psi_nond}
\end{equation}
Moreover we achieve the dimensionless form of EOSs for the GEOS as
\begin{equation}
p = \rho + \kappa \ln\,(\rho)
\label{eq:GEOS_nond},
\end{equation}
where $\kappa = \dfrac{\mathcal{P}_0}{c_{\rm s}^2 \rhoc}$ and $\kappa = 0$ gives the IEOS. The MPEOS dimensionless form is
\begin{equation}
p = 1 + A \ln\,(\rho)
\label{eq:MPEOS_nond},
\end{equation}
and finally the dimensionless form of the PEOS is
\begin{equation}
p = \rho^{1+1/n}
\label{eq:PEOS_nond}.
\end{equation}


\bsp	
\label{lastpage}
\end{document}